\documentclass{appolb}
\usepackage{graphicx}
 \usepackage{amsmath} 
 \usepackage[T1]{fontenc}
\usepackage{amssymb}
\usepackage{amsfonts}
\usepackage{cite}

\usepackage[colorlinks,citecolor=blue,linktoc=all,linkcolor=cyan]{hyperref}

\begin{document}
\title{Partial $A_4$ flavor symmetry of the leptonic 3HDM
\thanks{Presented by Joris Vergeest at the L International Conference of Theoretical Physics “Matter to the Deepest 2025”, Katowice, Poland}%
}
\author{Bartosz Dziewit, Marek Zra\l ek, Joris Vergeest
\address{Silesian University, Katowice, Poland}
}
\maketitle
\begin{abstract}
When the Higgs doublets in the 3HDM transform as a flavor triplet of the $A_4$ group, the lepton mass matrices accommodate the experimental neutrino mixing angles at arbitrary precision while maintaining the correct mass ordering of the charged and neutral leptons, the latter being Dirac neutrinos in the normal spectrum. Under $A_4$ symmetry, also agreement of the lepton masses with experimental data is obtained for Higgs vacua differing from those for fitting $U_{\text{PMNS}}$. For groups of order equal or less than 600 no contractions different from the one found for $A_4$ yield better agreement with experimental data, and the solution structure presented is unique within the set of groups studied.
\end{abstract}

\section{Introduction}
The flavor symmetry $F = U(3)^5$, inherent to the Standard Model, must be broken down to $U(1)$ below the electro-weak scale, should the fermion mass matrices be realistic, given the present phenomenological bounds. The lepton sector of the SM additionally suffers from vanishing masses of its neutrinos, which can only be avoided at the cost of a lepton number breaking term. If the existence of right-handed neutrinos is postulated ($F = U(3)^6$), neutrinos acquire mass by the Dirac mechanism. In this scenario the mass matrices of the neutrinos and the charged leptons would encode the masses and mixing of the leptons at energy below the EW scale. However, the Dirac or Majorana nature of neutrinos is as yet not decided and it is still an open question how many and which scalar fields contribute to the mass generation, thus leaving the Yukawa couplings as free parameters.
As an attempt to reduce the number of free parameters, and to find some reason for the structures of the fermion sectors, it has been proposed to replace $F$, somewhere above the EW scale, by a finite discrete symmetry $G \subset F$, commuting with the gauge sector. $F$ may subsequently further be broken down, differently for different fermion sectors.
This approach proved successful to achieve accordance with CKM and/or PMNS data \cite{Yao:2015dwa, Jur_iukonis_2017} for a number of extensions to the SM. However, the disparate mass hierarchies of the leptons compared to the quarks remain notoriously hard to be protected by $G$. This also holds specifically for the mass proportions of the charged leptons versus those of the neutrinos.

Within the lepton sector the scope of this work is a modestly expanded SM: the Higgs sector is extended with two Higgs doublets, thus forming a 3-Higgs-doublet model (3HDM). As we admit Dirac neutrinos, three right-handed neutrino fields are included as well.
We make no further assumptions about the presence of scalar flavons, invisible fields or heavy neutrinos, on which a flavor symmetry could be imposed. It has been shown that flavor symmetries of the 2HDM lead to lepton masses and mixing incompatible with experimental data \cite{Chaber:2018cbi}, which was one of the motivations to study the 3HDM.
Crucial in our approach is that not only the left-handed doublets and the right-handed singlets will transform as a flavor representation $\bf{3}$ of $G$ but the three Higgs doublets will do so as well. As a consequence, the three VEVs do not have a fixed alignment but become free parameters. This freedom would violate the constraints from the 3HDM Higgs scalar sector, which imposes a rigid vacuum alignment that rules out any symmetry left in the fermion sector, up to trivial mixing and/or fully degenerate masses \cite{Darvishi_2021,Gonz_lez_Felipe_2014}.
However, under some assumptions for the 3HDM (e.g. that the right-handed Dirac neutrinos transform as flavor singlets under $S_4$), viable mass matrices can be obtained \cite{10.1093/ptep/ptad075,izawa2023s4leptonflavormodel}. In our study we insist on handling the doublets and singlets of the charged leptons and those of the neutrinos on equal footing in the 3HDM, and assume that the Dirac mechanism solely is responsible for mass generation. As a consequence the invariance equations for the charged leptons and those for the neutrinos are very similar, making it hard to accommodate the lepton masses and mixing in a $n$HDM for $n<3$ \cite{Chaber:2018cbi}.

Our primary motivation to ignore the constraints on the VEVs is that non-vanishing
Yukawa matrices solely could be responsible for the breakdown of $F$ into $G$.
In a recent study of the 3HDM for all finite subgroups of $U(3)$ of order up to 1032, it has proved impossible to accommodate both the charged lepton and neutrino mass hierarchies when contractions
${\bf 3} \times {\bf 3'} \times {\bf 3''} = n {\bf 1} + ...$
are taken for $n=1$ only \cite{DZIEWIT2024138667}. We will show that the correct proportions of the masses can be obtained with $n=2$ for both of the mass terms, and a viable $U_{\text{PMNS}}$ can be accommodated.

In Section 2 we define the leptonic mass terms of the 3HDM. The scope and methodology to identify a viable $G$-symmetry of the Yukawa sector is briefly sketched in Section 3. In Section 4 the results for groups of order |G| < 600 are presented and conclusions are drawn in Section 5.

\section{The leptonic 3HDM and its flavor triplets}
The relevant terms of the Yukawa Lagrangian $\mathcal{L}^D$, based on the Dirac mechanism, are
\begin{eqnarray}
\mathcal{L}^l  \sim  \overline{l}_L \; \phi^{0\star}_i h^l_i \; l_R & \! \! \!\rightarrow & \! \! \! M^l \sim  v_i^\star h^l_i \label{terml} \\
\mathcal{L}^{\nu}  \sim  \overline{\nu}_L \; \phi^0_i h^{\nu}_i \nu_R & \! \! \! \rightarrow & \! \! \! M^{\nu} \sim v_i h^{\nu}_i \label{term2}
\end{eqnarray}
(leaving out the H.c terms of the Lagrangians), with $i, j = 1, 2, 3$. The mass matrices $M^l$ and $M^{\nu}$ each are built as a sum of 3
Yukawa matrices $h_i^{l,\nu}$. The Higgs field $\phi^ 0_i$ is the electrically neutral component of the $i$th Higgs doublet. The VEVs $v_i$ satisfy $\sqrt{\Sigma |v_i|^2}=(\sqrt{2}G_F)^{-1/2}$.
$\mathcal{L}^D \equiv \mathcal{L}^l + \mathcal{L}^{\nu}$ is flavor symmetric under discrete group $G$ if and only if (not necessarily different) unitary matrix representations of $G$ can be assigned to the flavor triplets such that $\mathcal{L}^D$ remains invariant for some choice of non-vanishing Yukakawa matrices.

Particle experiments may involve the charged current term
\begin{eqnarray}
\mathcal{L}_{CC} \sim \overline{l}_L \nu_L = \overline{l}_{L \text{mass}} U^{l \dagger} U^{\nu} \nu_{L \text{mass}},
\end{eqnarray}
with unitary matrices $U^l$ and $U^{\nu}$ defining the (unitary) neutrino mixing matrix $U_{\text{PMNS}} = U^{l \dagger} U^{\nu}$. For postulated $M^l$ and $M^\nu$ the matrices $U^l$ and $U^{\nu}$ can be obtained by bi-diagonalization:
\begin{eqnarray}
\label{Ml} M^l & = & V^l M_{diag}^l U^{l \dagger} \label{Ml} \\
\label{Mnu} M^{\nu} & = & V^{\nu} M_{diag}^{\nu} U^{\nu \dagger}. 
\end{eqnarray}
The entries of the diagonal matrices are proportional to the mass values of the charged leptons and the neutrinos, respectively. It is important to note that, at this point, the entries of the diagonal matrix arise in arbitrary order.

\section{Mass matrices invariant under $G$}
Let for a given element of $G$ and for given choice of group representations, the three triplets of $\mathcal{L}^l$ transform by matrices $A$, $B$ and $C$, respectively. Then $\mathcal{L}^l$ is invariant under this transformation if and only if
\begin{eqnarray}
\overline{(Al_L)} (C^\star \phi^{0\star})_j  h^l_j \; B l_R \label{ABC} = \overline{l_L} \phi^{0\star}_i h^l_i  \; l_R 
\end{eqnarray}
for all triplets $l_L$, $l_R$, $\phi^{0}$. For given matrices $A$, $B$, $C$ the matrices $h^l_i$ then satisfy
\begin{eqnarray} \label{vec1}
(C^\dagger \otimes B^T \otimes A^\dagger) vec(h^l) = vec(h^l)   \label{kron01},
\end{eqnarray}
where $\otimes$ denotes the Kronecker product and $vec(h^l)$ is the column vector (of length 27) built from the column vectors of $h^l_i$ for $i = 1,2,3$ \cite{broxson2006}.
Before actually solving the systems of linear equations, the dimensionality of the $h$ spaces can be determined based on the characters of $G$.

The character table of a finite group $G$ lists its irreducible characters and facilitates the decomposition of tensor products of representations. If
\begin{eqnarray}
{\bf 3}_C^\star \times {\bf 3}_B \times {\bf 3}_A^\star = n^l {\bf 1} + ...
\end{eqnarray}
then Eq. \eqref{kron01} has $n^l$ inequivalent solutions. It implies that $n^l$ different contractions lead to the invariant flavor singlet $\textbf{1}$ of $G$, and hence to a $G$-invariant $\mathcal{L}^l$. 
Let $G$ have $f$ generators and ${\bf 3}_A$, ${\bf 3}_B$ and ${\bf 3}_C$ be a choice of representations. Then a system of $f$ linear equations as given in Eq. \eqref{kron01} is specified, yielding $n^l$ inequivalent solutions $h^l$, where $0 \le n^l \le 3$ . For each of these solutions and for any choice of VEVs $v_i$ the mass matrix $M^l$ can be calculated up to an overall complex constant, using Eq. \eqref{Ml}. $h^{\nu}$ and $M^{\nu}$ are obtained similarly, using representations ${\bf 3}_A$, ${\bf 3}_D$ and ${\bf 3}_C$, where ${\bf 3}_D$ transforms triplet $\nu_R$.

\section{The mass matrices generated for $|G| \le 600$ }
Even when $G$ is not a subgroup of $U(3)$ it may generate relevant sets of representations. The selection criterion is that a candidate $G$ has at least one 3D irrep, which may or may not be faithful. Clearly, an unfaithful ${\bf 3}_i$ of $G$ will be isomorphic to a faithful ${\bf 3}_j$ of some smaller group, but could nonetheless be contained in a set $\{{\bf 3}_A$, ${\bf 3}_B$, ${\bf 3}_C$, ${\bf 3}_D\}$ that produces solutions $h^l$ and $h^{\nu}$, which would not be feasible using faithful ${\bf 3}$s only. Of the 3221 accepted groups with $|G| \le 600$, 487 are a subgroup of $U(3)$. 

$n^l$ and $n^{\nu}$ determine the dimensionality of the spaces of invariant mass matrices. The 10 observables of the postulated 3HDM lepton sector are the 6 fermion masses, 3 PMNS mixing angles, and 1 CP phase. Since we do not expect to obtain absolute mass scales from group theory, we can derive at most 2 charged lepton mass ratios, 2 neutrino mass ratios, and the 4 PMNS parameters given a $G$-invariant set \{$M^l, M^{\nu}$\}. The $v_i$ are free parameters up to a common complex constant. So we have 8 quantities to be fitted with 4 (real) parameters ($v_2/v_1$ and $v_3/v_1$), given any $G$-symmetry of $\mathcal{L}^D$. Taking into account the degrees of freedom of the $h^l_i$ and $h^{\nu}_i$ spaces, the number of free parameters is $2(n^l + n^{\nu})$, ranging from 4 (when $n^l = n^{\nu} = 1)$ to 12 (when $n^l = n^{\nu} = 3)$.

\begin{table}
\centering
\begin{tabular}{c | c c c}
$n^l \backslash n^{\nu}$ & 1 &2 &3 \\
\hline
1 &  20,437 & 3816 & 0\\
2 & 3816 &  729 & 0\\
3 &  0   &  0 & 0 \\
\hline
\end{tabular}
\caption{\label{tab:dim} Number of $G$-invariant solutions of dimensions $n^l$ and $n^{\nu}$}
\end{table}

The scan of groups yields 28,807 inequivalent $G$-invariant pairs \{$M^l, M^{\nu}$\}. The occurring dimensionalities of the latter are shown in Table \ref{tab:dim}.
An analysis of the solutions with $n^l=n^{\nu}=1$ revealed that none of them can give realistic lepton masses \cite{DZIEWIT2024138667}.
We find that only solutions with $n^l = n^{\nu} = 2$ can lead to viable mixing angles, so we focus on them from here on. (As a side note, 3-dimensional solutions do exist for $\mathcal{L}^l$ and $\mathcal{L}^{\nu}$ separately, but not for $\mathcal{L}^D$).

The mass matrices $M^l$ and $M^{\nu}$ encountered for the selected groups can be partitioned into 7 subsets based on the matrix pattern. In the present study it turns out that one specific pattern is needed: a monomial matrix of which the nonzero entry in row $i$ is proportional to $v_j$, $j= $ mod($i+k,3)+1$ for some $k \in \{0,1,2)$.

\section{Lepton mass hierarchy}
\label{leptonmass}
A given $G$-invariant pair $\{M^l,M^{\nu}\}$ defines constraints on the lepton masses and neutrino mixing of the 3HDM theory. If the constraints are in accordance with the experimental observations, it could support the hypothesis of the flavor-breaking mechanism induced by $G$. In the analysis we check whether the experimentally known masses of the charged leptons and of the neutrinos can be reproduced by the given pair of mass matrices. Further, we verify the neutrino mixing angles. In the latter step we must take $U_{\text{PMNS}}$ consistent with the ordering of the masses; the rows and columns of the calculated $U_{\text{PMNS}}$ cannot be freely interchanged but are fixed by ordering the eigenvalues increasingly in Eqs. \eqref{Ml},\eqref{Mnu}.

\begin{figure*}
\includegraphics[scale=1.0]{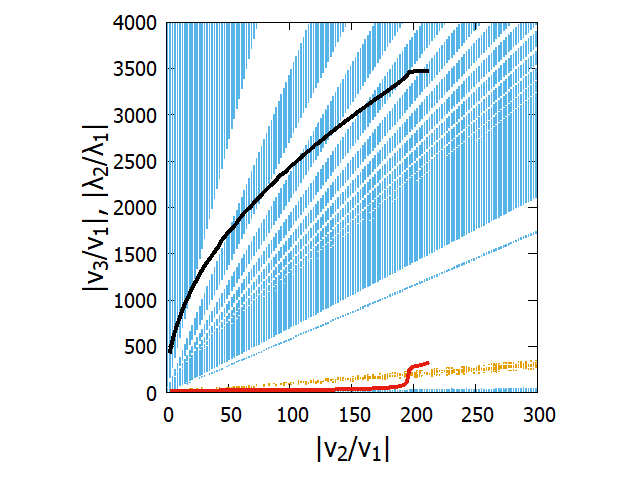}
\caption{Case $n^l = n^{\nu} = 2$. The blue (yellow) points are projections onto the $|v_2/v_1|-|v_3/v_1|$ plane from calculations of NO (IO) neutrino mass ratios consistent with experimental data. The black curve represents calculations of correct charged lepton mass ratios. The intersection of the latter curve and the blue area signifies the calculations of viable lepton masses. The red curve displays $|\lambda^l_2/\lambda^l_1|$ as function of $|v_2/v_1|$ from the calculations that produce the black curve.}
\label{fig:dmix}
\end{figure*}

\begin{figure*}
\includegraphics[scale=1.0]{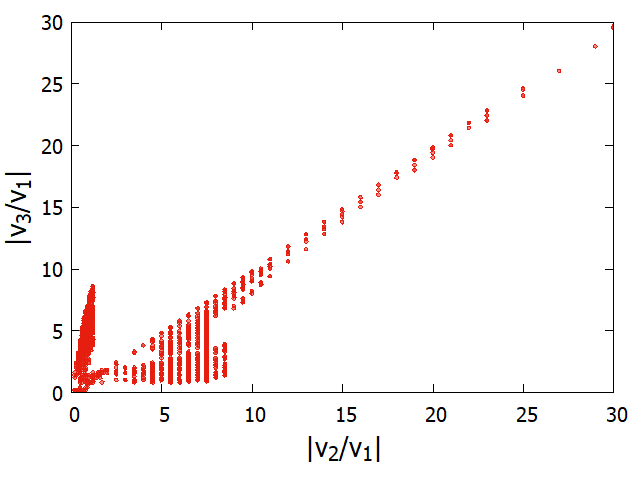}
\caption{Case $n^l = n^{\nu} = 2$. Points projected onto the $|v_2/v_1|-|v_3/v_1|$ plane yielding PMNS mixing angles consistent with experimental data, satisfying $\chi^2 (\sin^2 \theta_{12}) + \chi^2 (\sin^2 \theta_{23}) + \chi^2 (\sin^2 \theta_{13}) < 3$.}
\label{fig:PMNS1}
\end{figure*}

There are 729 inequivalent joint solutions for $n^l = n^{\nu} = 2$. These can be obtained cumulatively from group $A_4$ and three more groups, with GAP-ID [36,3], [108,3] and [324,3] \cite{GAP4}. The solutions reappear in groups of higher order.
All solutions lead to the same structure of $M^l$ and $M^\nu$:
\begin{eqnarray}
\label{P1P1-P1P1form1}
m^l & \sim &
\begin{pmatrix}
 0  &  \lambda^l_2 v_3  &  \lambda^l_1 v_2  \\    
  \lambda^l_1 v_3 &  0   &  \lambda^l_2 v_1 \\
 \lambda^l_2 v_2  & \lambda^l_1 v_1  &  0  \\
\end{pmatrix}\\
m^{\nu} & \sim &
\begin{pmatrix}
 0  &  \lambda^{\nu}_2 v_3  &  \lambda^{\nu}_1 v_2  \\    
  \lambda^{\nu}_1 v_3 &  0   &  \lambda^{\nu}_2 v_1 \\ 
 \lambda^{\nu}_2 v_2  & \lambda^{\nu}_1 v_1   &  0  \\
\end{pmatrix},
\end{eqnarray}
where it is understood that each occurrence of $v_i$ carries an additional phase, not shown.

For given $|v_2/v_1| \in [0, m_{\mu}/m_e \approx 206.6]$ we find unique $|v_3/v_1|$ and $|\lambda^l_2/\lambda^l_1|$ for which $M^l$ gives the correct charged lepton mass ratios, depicted by the black and red curves in Fig. \ref{fig:dmix}.
Furthermore we obtain viable neutrino mass ratios for several continua of $|v_3/v_1|$, represented by the blue areas in Fig. \ref{fig:dmix} for NO (normal ordered) neutrinos. The narrow yellow band in the lower part of the figure represents the IO (inverted order) neutrino fits.
The intersection of the black charged leptons curve with the blue shaded areas defines the quantities $v_i, \lambda^l_i$ and $\lambda^{\nu}_i$ leading to calculated mass ratios that are compatible with the experimental data of all leptons.
If it is assumed that neutrinos are normally ordered we find $R_{\text{NO}} := \frac{\Delta m^2_{21}}{\Delta m^2_{32}} \approx 0.0307125$, a small fraction of $\sigma$ away from its measured value.
Given the calculated neutrino mass ratios and the measured value of $\Delta m^2_{21}$ the absolute scale of the neutrino masses can be calculated as
\begin{eqnarray}
\label{m1}
m_1 &\approx & 5.0 \times 10^{-6} \>  \text{eV}\\
m_2 & \approx & 0.86 \times 10^{-2} \>  \text{eV}\\ 
\label{m2}
m_3 & \approx & 5.03  \times 10^{-2} \>  \text{eV}.
\end{eqnarray}
These values are well within all experimental bounds \cite{ParticleDataGroup:2024cfk}.
Whereas $|\lambda^l_2/\lambda^l_1|$ (red curve in Fig. \ref{fig:dmix}) of the viable solutions increases from about 20 to 100 for $|v_2/v_1|$ from 0 to 206.6, the quantity $|\lambda^{\nu}_2/\lambda^{\nu}_1|$ rises slowly from 5.8 to 6.2 (not shown). The calculated neutrino masses are the same at all points on the black curve.
We remark that the lepton mass calculations just described are also obtained with $n^l$=1 and $n^{\nu}=2$ under flavor symmetry group $Z_7 \rtimes Z_3$, GAP-ID [21,21] \cite{vergeest3hdm}.

\section{Neutrino mixing}
The calculated $\sin^2\theta_{12}, \sin^2\theta_{23}, \sin^2\theta_{13}$ depend on 7 real parameters: $|v_2/v_1|, v_3/v_1, \lambda^l_2/ \lambda^l_1$ and $\lambda^{\nu}_2/\lambda^{\nu}_1$, where the phase of $v_2$ is absorbed by the phases of $\lambda^l_2$ and $\lambda^{\nu}_2$. Since the parameters are unbounded the search for $\theta_{ij}$ that are near the observed values tends to be incomplete. The projection onto the $(|v_2/v_1|, |v_3/v_1|)$ plane of the parameter regions where
\begin{eqnarray}
\chi^2 (\sin^2 \theta_{12}) + \chi^2 (\sin^2 \theta_{23}) + \chi^2 (\sin^2 \theta_{13}) < 3
\end{eqnarray}
are shown in Fig. \ref{fig:PMNS1}. The mass matrices giving these results can be chosen such that $m_e < m_{\mu} < m_{\tau}$ and $m_1 < m_2 < m_3$. However, the calculated mass values themselves are not compatible with experimental data. The regions depicted in Fig. \ref{fig:PMNS1} are far below the blue zone of Fig. \ref{fig:dmix} and they are also significantly below the charged-leptons curve in the latter figure. For  given $|v_2/v_1|  \gtrsim 10$ the viable mixing angles appear restricted to $|v_3/v_1| \in [|v_2/v_1| - 2, |v_2/v_1|]$. The solutions compatible with the neutrino mixing angles produce $\delta _{CP}$ in the range of 1.5 to 1.9 radians, which are not in accordance with the experimentally obtained results of $-1.9 \pm 0.6$ radians.

\begin{table}
\begin{tabular}{l| c c c  c c c c }
 & $n^l$ & $n^{\nu}$ & mass     & mass & mass  & mass & $G_F$ \\
 & & & order   &  ratios &  order &  ratios  \\
 & & &         &         & +PMNS  & +PMNS\\
\hline
1HDM & 1 & 1 & {\Large \color{red} x} & {\Large \color{red} x } & {\Large  \color{red} x } & 
  {\Large \color{red} x }\\
\hline
2HDM  & 1 & 1 & {\Large \textbf{ \color{green} \checkmark}} & {\Large \color{red} x } & {\Large \color{red} x } & 
  {\Large \color{red} x } & $S_4$\\
\hline
3HDM & 1 & 1 &  {\Large \textbf{ \color{green} \checkmark}}  & {\Large \color{red} x } & {\Large  \color{red} x } & 
{\Large \color{red} x } & $Z_7 \rtimes Z_3$\\ 
& 1 & 2 & {\Large \textbf{ \color{green} \checkmark}} & {\Large \textbf{ \color{green} \checkmark}} & {\Large \color{red} x }
& {\Large \color{red} x } & $Z_7 \rtimes Z_3$\\
& 2 & 2 & {\Large \textbf{ \color{green} \checkmark}}& {\Large \textbf{ \color{green} \checkmark}} & 
{\Large \textbf{ \color{green} \checkmark}} & 
{\Large \color{red} x } & $A_4$\\
\hline
\end{tabular}
\caption{\label{tab:nHDM} $G_F$ is the smallest flavor symmetry group that has a 3D irrep and leads to realistic masses and/or mixing, as indicated by the check marks for given $n^l$ and $n^{\nu}$, obtained for the lepton sector of the nHDM, assuming the neutrinos are Dirac particles.}
\end{table}

\section{Conclusions}
Table \ref{tab:nHDM} summarizes for which minimal nontrivial $G$ and to which extent the leptonic nHDM up to $n=3$ is flavor symmetric when the neutrinos are Dirac particles.
Viable neutrino mixing angles are obtained with the 3HDM when both $M^l$ and $M^{\nu}$ originate from two contractions in flavor space, as realized with group $A_4$. $U_{\text{PMNS}}$ following from this symmetry has its rows and columns fixed, consistent with the ordering of the masses of the charged leptons and the neutrinos in normal ordering. Although the mass orderings can be accommodated, the experimental masses themselves cannot. The solutions imply $\delta_{CP} \in$ [1.5, 1.9] radians, not in accordance with -1.9 $\pm$ 0.6 radians from experiments. These results are obtained under the assumption that the $SU(2)$ Higgs doublets transform as a triplet in flavor space and thus avail the VEVs as free parameters. Several groups with order higher than 12 generate mass matrices that partially fit the experimental data in exactly the way as did $A_4$. The solution obtained from $A_4$ appears to be unique for all selected groups of order 600 or less.

Groups leading to $n^{l,\nu}=3$ different contractions to build the mass matrices do occur, but the requirements on the representations for the 3HDM, described in Section 3, are not fulfilled. So it can be concluded that for $n^l=n^{\nu}=2$, $A_4$ admits the maximal viability of the proposed 3HDM attainable under non-trivial discrete symmetry of order 600 or less.

In a follow-up study the three-dimensional representations that are not irreducible could be taken into account, although the number of possible ${\bf1'+\bf1''+\bf1'''}$ representations explode combinatorially with $|G|$.
The direct sums ${\bf2 + \bf1'}$ might be relevant. In the 3HDM the effective Majorana neutrino term leads to new invariance equations and hence to neutrino mass matrices that may be inequivalent with all mass matrices found in this work.

\section*{Acknowledgments}
This work has been supported in part by the Polish National Science Center (NCN) under grant 2020/37/B/ST2/02371 and the Research Excellence Initiative of the University of Silesia in Katowice, Poland.

\bibliographystyle{elsarticle-num}
\bibliography{bibliography.bib}

\end{document}